# Sine function with a cosine attitude

A. D. Alhaidari

*Shura Council, Riyadh 11212, Saudi Arabia*
AND
*Physics Department, King Fahd University of Petroleum & Minerals, Dhahran 31261, Saudi Arabia*
E-mail: haidari@mailaps.org

We give a revealing expose that addresses an important issue in scattering theory of how to construct two asymptotically sinusoidal solutions of the wave equation with a phase shift using the same basis having the same boundary conditions at the origin. Analytic series representations of these solutions are obtained. In 1D, one of the solutions is an even function that behaves asymptotically as sin(*x*), whereas the other is an odd function, which is asymptotically cos(*x*). The latter vanishes at the origin whereas the derivative of the former becomes zero there. Eliminating the lowest *N* terms of the series makes these functions vanishingly small in an interval around the origin whose size increases with *N*. We employ the tools of the J-matrix method of scattering in the construction of these solutions in one and three dimensions.

## 1. Introduction:

To study the structure and dynamics of a subatomic system, physicists perform scattering experiments. In such an experiment, a uniform flux of particles (referred to as probes or projectiles) will be incident on the target under study. The scattered flux will carry the needed information about the system. In potential scattering theories [1], a configuration potential function that models the structure of the system and its interaction with the probes is proposed. This model will be tested against the outcome of the experiment. Typically, such a potential function has a finite range such that in the asymptotic region, where the incident particles are ejected and then collected, its value is zero. Therefore, the incident and scattered particles are free and, thus, represented by sinusoidal wavefunctions. The information about the system is contained in the difference between the two angles ("phase shift") of these sinusoidal functions representing the incident and scattered flux. Mathematically, this means that one needs two independent sinusoidal solutions of the free wave equation such as $\sin(kx)$ and $\cos(kx+\delta)$, where $x$ is the relevant configuration space coordinate and $k$ is the wave number which is related to the incident beam energy as $k=\sqrt{2E}$. The phase shift $\delta$ depends on the energy, angular momentum, and model potential parameters [2].

Theoretically, the scattering problem at steady state is described by the solution of the time-independent energy eigenvalue equation $(H-E)|\chi\rangle=0$, where $H$ is the Hamiltonian operator and the energy $E$ is positive and continuous. For a general dynamical system, the analytic solution of this equation is often very difficult to obtain.

However, for a large class of problems that model realistic physical systems, the Hamiltonian could be written as the sum of two components: $H = H_0 + V$. The "reference" Hamiltonian $H_0$ is often simpler and carries a high degree of symmetry. It is treated analytically despite its infinite range. The "potential" $V$ that models the system is not, but it is usually endowed with either one of two properties. Its contribution is either very small compared to $H_0$ or is limited to a finite region in configuration or function space. Perturbation techniques are used to give a numerical evaluation of its contribution in the former case, whereas algebraic methods are used in the latter. As for the scattering problem under consideration, we will only be concerned with the latter case where the potential is assumed to vanish beyond a certain finite range. Thus, the analytic problem is confined to the solution of the reference wave equation, $(H_0 - E)|\psi\rangle = 0$. As discussed above, the solution to this problem is of fundamental importance in scattering since it will be the carrier of information about the structure and dynamics of the system. Typically, there are two independent solutions of this problem. Both are sinusoidal since they represent free particle motion due to the fact that in the asymptotic region $V = 0$. The phase difference between the two sinusoidal solutions will contain the scattering information. To obtain a full solution of the scattering problem, one needs (beside the kinematical solution of the reference problem) a numerical evaluation of the dynamics contained in the contribution of the potential $V$. In the algebraic methods of scattering, such an evaluation is carried out by computing the matrix elements of these operators in a complete set of square integrable basis. Conditions at the interface between the scattering and asymptotic regions dictate that the same basis (or linear combinations thereof) be used for both asymptotic solutions. This implies that these two independent asymptotic solutions (the sine and cosine) will have the same symmetry attributed to the basis of the solution space and satisfy the same boundary conditions at the origin. For example, we shall see that in 1D the cosine solution is required to be an odd function that vanishes at the origin whereas the sine solution is an even function with a zero derivative at the origin. Moreover, looking at these functions far away from the origin, one could never predict correctly this curious and non-trivial behavior. Mathematically, it is intriguing to find an analytic realization of such functions. Physically, of course, this is fully acceptable since these solutions have relevance only outside the scattering region, and whatever happens to them near the origin is irrelevant. In fact, they are even allowed to diverge there.

In this work, we use the J-matrix theory of scattering [3,4], as an example of an algebraic method, in the construction of these interesting reference solutions. In the following section, we find even and odd solution spaces for the reference wave equation in one dimension and demonstrate how to construct an even (odd) function that behaves as sine (cosine) with cosine-like (sine-like) properties at the origin. These will be given as convergent series involving products of Hermite polynomials and confluent hypergeometric functions. In the Appendix, we also construct the asymptotic sinusoidal solutions of the scattering problem in three dimensions having these interesting properties. Those will be given as series of products of Laguerre polynomials and hypergeometric functions. To illustrate the captivating and intriguing properties of these functions, graphical representations will be given. We will also show that by eliminating the lowest $N$ terms of the series, these functions become vanishingly small in an interval around the origin whose size increases with $N$.

## 2. The one-dimensional case:

One-dimensional nonrelativistic steady flux scattering is represented by the solution of the following time-independent Schrödinger equation

$$\left[-\frac{\hbar^2}{2m}\frac{d^2}{dx^2} + V(x)\right]\chi(x,E) = E\,\chi(x,E), \quad x \in \mathbb{R}, \tag{1}$$

where $V(x)$ is the scattering potential that models the structure of the target and its interaction with the probes. It is assumed to have a finite range but is not required to be analytic. Thus, the solutions of the reference problem, where $V = 0$, are also the asymptotic solutions of Eq. (1) and carry the scattering information. If we let $\psi(x,E)$ stand for the reference wavefunction, then in the atomic units $\hbar = m = 1$ we can write $\left(\frac{d^2}{dx^2} + k^2\right)\psi(x,E) = 0$. Measuring length in units of $\lambda^{-1}$ (e.g., the Bohr radius), where $\lambda$ is a positive parameter, we can write this equation in terms of dimensionless quantities as

$$\left(\frac{d^2}{dy^2} + \mu^2\right)\psi(y,\mu) = 0, \qquad y \in \mathbb{R}, \tag{2}$$

where $y = \lambda x$ and $\mu = k/\lambda$. Consequently, the two independent real solutions of this reference wave equation are $f_\mu^+(y) = A\cos(\mu y)$ and $f_\mu^-(y) = B\sin(\mu y)$, where $A$ and $B$ are arbitrary real constants. Due to space reflection symmetry of the reference wave operator, $J(\mu) = \frac{d^2}{dy^2} + \mu^2$, parity is conserved and the solution space of the reference problem splits into two disconnected subspaces; even and odd. $f_\mu^+$ ($f_\mu^-$) belongs to the even (odd) subspace. Therefore, to perform the algebraic calculations, we need two independent square integrable bases sets that are orthogonal to each other. They must be compatible with the two sub-domains of the reference Hamiltonian, which are invariant eigen-spaces of the parity operator. In general, the model potential $V(x)$ does not have a definite parity. Therefore, both bases will be used in the calculation of the scattering phase shift. Now, such complete bases that are compatible with the two sub-domains of $H_0$ and with the 1D problem have the following elements

$$\phi_n^+(y) = \mathcal{A}_n^+ e^{-y^2/2} H_{2n}(y), \tag{3a}$$

$$\phi_n^-(y) = \mathcal{A}_n^- e^{-y^2/2} H_{2n+1}(y), \tag{3b}$$

where $n = 0, 1, 2, \ldots$ and $H_m(y)$ is the Hermite polynomial of order $m$ [5]. The normalization constants are taken as $\mathcal{A}_n^+ = \frac{1}{\sqrt{\sqrt{\pi}\,2^{2n}(2n)!}}$ and $\mathcal{A}_n^- = \frac{1}{\sqrt{\sqrt{\pi}\,2^{2n+1}(2n+1)!}}$. Using the orthogonality relation of the Hermite polynomials, one can easily verify that these two bases are orthogonal to each other (i.e., $\langle\phi_n^+|\phi_m^-\rangle = 0$) and that the elements of each one of them form an orthonormal set (i.e., $\langle\phi_n^\pm|\phi_m^\pm\rangle = \delta_{nm}$). Moreover, they are related as $\phi_n^-(x) = \phi_{n+1/2}^+(x)$. The integration measure is $\int_{-\infty}^{+\infty} \ldots dy$. Similar to $f_\mu^+(y)$, the elements $\{\phi_n^+\}_{n=0}^\infty$ are even in $y$ and satisfy the same condition at the origin; namely, $\left(df_\mu^+/dy\right)_{y=0} = 0$. On the other hand, $\{\phi_n^-\}_{n=0}^\infty$ are odd in $y$ like $f_\mu^-(y)$ and satisfy the same condition at the origin; namely, $f_\mu^-(0) = 0$.

Now, the reference solutions $f_\mu^\pm(y)$ belong to the two linear subspaces spanned by $\{\phi_n^\pm(y)\}_{n=0}^\infty$, respectively. In other words, they are expandable in these two bases sets as the following infinite sums (Fourier-like expansion)

$$f^+(y) = A\cos(\mu y) = \sum_{n=0}^\infty s_n^+(\mu)\phi_n^+(y), \qquad (4a)$$

$$f^-(y) = B\sin(\mu y) = \sum_{n=0}^\infty s_n^-(\mu)\phi_n^-(y), \qquad (4b)$$

where $s_n^\pm$ are the expansion coefficients that are functions of the energy parameter $\mu$. Due to the completeness of the basis and that they are square integrable, one can prove that the two series (4a,4b) are bounded and convergent for all $y$ and $\mu$. Nonetheless, this will also be demonstrated numerically in the results and graphs obtained below. Now, to calculate the expansion coefficients, we multiply both sides of Eq. (4a) and Eq. (4b) by $e^{-y^2/2}H_{2m}(y)$ and $e^{-y^2/2}H_{2m+1}(y)$, respectively. Integrating both sides and using the orthogonality relation of the Hermite polynomials and the integral formulas [6],

$$\int_{-\infty}^{+\infty} e^{-y^2/2}\cos(\mu y)H_{2n}(y)dy = (-1)^n\sqrt{2\pi}\,e^{-\mu^2/2}H_{2n}(\mu), \qquad (5a)$$

$$\int_{-\infty}^{+\infty} e^{-y^2/2}\sin(\mu y)H_{2n+1}(y)dy = (-1)^n\sqrt{2\pi}\,e^{-\mu^2/2}H_{2n+1}(\mu), \qquad (5b)$$

we obtain

$$s_n^+(\mu) = \frac{(-1)^n \pi^{1/4}\sqrt{2}A}{2^n\sqrt{\Gamma(2n+1)}} e^{-\mu^2/2}H_{2n}(\mu), \text{ and} \qquad (6a)$$

$$s_n^-(\mu) = \frac{(-1)^n \pi^{1/4}B}{2^n\sqrt{\Gamma(2n+2)}} e^{-\mu^2/2}H_{2n+1}(\mu). \qquad (6b)$$

These are also related as follows: $s_n^-(E) = -i\frac{B}{A}s_{n+\frac{1}{2}}^+(E)$. Using the differential equation, recursion relation, and orthogonality property of the Hermite polynomials [5], we obtain the following tridiagonal matrix elements of the reference wave operator

$$\left\langle \phi_n^\pm \left| \left(\frac{d^2}{dy^2} + \mu^2\right) \right| \phi_m^\pm \right\rangle \equiv J_{nm}^\pm(\mu)$$
$$= \left[\mu^2 - \left(2n+1\mp\tfrac{1}{2}\right)\right]\delta_{n,m} + \sqrt{n\left(n\mp\tfrac{1}{2}\right)}\,\delta_{n,m+1} + \sqrt{(n+1)\left(n+1\mp\tfrac{1}{2}\right)}\,\delta_{n,m-1} \qquad (7)$$

Therefore, the matrix representation of the reference wave equation (2) becomes equivalent to the following three-term recursion relation for the expansion coefficients

$$\mu^2 s_n^\pm = \left(2n+1\mp\tfrac{1}{2}\right)s_n^\pm - \sqrt{n\left(n\mp\tfrac{1}{2}\right)}\,s_{n-1}^\pm - \sqrt{(n+1)\left(n+1\mp\tfrac{1}{2}\right)}\,s_{n+1}^\pm, \qquad (8)$$

where $n = 1,2,3,...$. The initial relation ($n = 0$) is

$$\mu^2 s_0^\pm = \left(1\mp\tfrac{1}{2}\right)s_0^\pm - \sqrt{1\mp\tfrac{1}{2}}\,s_{n+1}^\pm. \qquad (9)$$

Therefore, we can in fact determine all of the expansion coefficients $\{s_n^\pm\}_{n=0}^\infty$ using only $s_0^\pm(\mu)$ as seed in the three-term recursion (8) and its initial relation (9). These seeds are

$$s_0^+(\mu) = \sqrt{2}A\pi^{1/4}e^{-\mu^2/2}, \quad s_0^-(\mu) = 2\pi^{1/4}B\mu e^{-\mu^2/2}. \qquad (10)$$

Moreover, using the differential equation of the Hermite polynomials one can easily verify that $s_n^\pm(\mu)$ satisfy the following second order differential equation in the energy

$$\left[\frac{d^2}{d\mu^2} - \mu^2 + 2(2n+1)\mp 1\right]s_n^\pm(\mu) = 0. \qquad (11)$$

Now, we are in a position to search for the "complementary" solutions to $f_\mu^\pm(y)$. As stated in the Introduction section, these solutions must be asymptotically sinusoidal with a phase that carries the scattering information as deviations from the corresponding phases of $f_\mu^\pm$. These complementary solutions, denoted as $g_\mu^\pm(y)$, must also belong to the same sub-domains of the reference Hamiltonian as does $f_\mu^\pm(y)$. That is, we should be able to write them as $g_\mu^\pm(y) = \sum_n c_n^\pm(\mu)\phi_n^\pm(y)$, where $c_n^\pm(\mu)$ are new expansion coefficients that are independent from $s_n^\pm(\mu)$. In the J-matrix method, these new expansion coefficients are required to satisfy the following criteria [4]:

(1) Be an independent solution of the 2$^{nd}$ order energy differential equation (11),
(2) Satisfy the same recursion relation (8) but not the initial relation (9), and
(3) Make the "complementary" reference wavefunctions $g_\mu^\pm(y)$ asymptotically sinusoidal and identical to $f_\mu^\pm(y)$ but with a phase difference that depends on the scattering potential parameters and the energy. In the absence of the potential, this phase difference is $\frac{\pi}{2}$.

The first condition is satisfied by the following independent solutions of Eq. (11):

$$c_n^+(\mu) = A D_n^+ \mu e^{-\mu^2/2} {}_1F_1\left(-n+\tfrac{1}{2}; \tfrac{3}{2}; \mu^2\right), \tag{12a}$$

$$c_n^-(\mu) = B D_n^- e^{-\mu^2/2} {}_1F_1\left(-n-\tfrac{1}{2}; \tfrac{1}{2}; \mu^2\right), \tag{12b}$$

where ${}_1F_1(a;c;z)$ is the confluent hypergeometric function and $D_n^\pm$ are normalization constants, which are determined by the second condition as

$$D_n^\pm = d^\pm \sqrt{\Gamma(n+1)/\Gamma\left(n+1\mp\tfrac{1}{2}\right)}, \tag{13}$$

where $d^\pm$ are constants; independent of $\mu$ and $n$. They are determined by the third condition as $d^+ = 2\sqrt{2}$ and $d^- = \sqrt{2}$. This could also be verified analytically and/or numerically. We also find that the $\lim_{|y|\to\infty} g_\mu^+(y) = A\sin(\mu y)$ and $\lim_{|y|\to\infty} g_\mu^-(y) = B\cos(\mu y)$. The initial recursion relations satisfied by $c_n^\pm(\mu)$ are not homogeneous like those of $s_n^\pm$ in Eq. (9). They contain source terms and read as follows

$$\left(\mu^2 - \tfrac{1}{2}\right)c_0^+ + \sqrt{\tfrac{1}{2}}\, c_1^+ = \pi^{-\frac{1}{4}}\sqrt{2} A\mu e^{\mu^2/2}, \tag{14a}$$

$$\left(\mu^2 - \tfrac{3}{2}\right)c_0^- + \sqrt{\tfrac{3}{2}}\, c_1^- = -\pi^{-\frac{1}{4}} B e^{\mu^2/2}. \tag{14b}$$

Thus, one needs only $c_0^\pm(\mu)$ as seeds in the three-term recursion relation (8) and initial relations (14) to obtain all $c_n^\pm(\mu)$ recursively. These seeds are

$$c_0^+(\mu) = 2\sqrt{2}\,\pi^{-\frac{1}{4}} A\mu e^{-\mu^2/2} {}_1F_1\left(\tfrac{1}{2}; \tfrac{3}{2}; \mu^2\right), \tag{15a}$$

$$c_0^-(\mu) = 2\pi^{-\frac{1}{4}} B e^{-\mu^2/2} {}_1F_1\left(-\tfrac{1}{2}; \tfrac{1}{2}; \mu^2\right). \tag{15b}$$

Now, we give visual illustrations of the functions $g_\mu^\pm(y)$ showing their intriguing properties. For $\mu = 1.2$, Fig. 1a shows $f_\mu^+(y)$ as a solid red curve and $g_\mu^+(y)$ as the solid blue curve. Nothing curious, of course, about the $f_\mu^+(y)$ curve; it is just the cosine function. However, it is interesting to observe that despite the fact that $g_\mu^+(y)$ is an even function of $y$, it behaves exactly as a sine function away from the origin and manages to deform itself near the origin to comply with the cosine-like boundary condition there (i.e., its derivative vanishes). On the other hand, Fig. 1b shows a similar curious behavior for the odd function $g_\mu^-(y)$ that behaves as a cosine function far from the origin but vanishes at the origin. Looking at these functions far from the origin, one may never be able to predict correctly these non-trivial behaviors near the origin. Another interesting property of the series representation obtained above for the sinusoidal functions $f_\mu^\pm(y)$ and $g_\mu^\pm(y)$ goes as follows. If the series were to be started not from $n = 0$ but from $n = N$, for some large enough integer $N$, then the resulting functions maintain their sinusoidal behavior outside an interval, which is symmetric around the origin and in which the functions become vanishingly small. The size of the interval increases with $N$. In fact, for any given range of the scattering potential $V(y)$, one can choose an $N$ that makes these functions vanishingly small within that range. As an example, we plot the even functions ($f_\mu^+$ and $g_\mu^+$) shown in Fig. 2a and Fig. 2b for $N = 10$ and $N = 20$, respectively. In the Appendix, we carry out the same development in three dimensions. The corresponding results are shown graphically in Fig. 3 and Fig. 4. Given a specific potential model $V$, scattering will introduce the phase shift angle $\delta$ into the sinusoidal functions $g_\mu^\pm(y)$.

## 3. Conclusion:

In one dimension, we were able to write the sine and cosine functions as infinite convergent series of products of odd and even Hermite polynomials. Complementary functions, which are asymptotically sinusoidal and required by the scattering theory, were also constructed as series of products of the Hermite polynomials and the confluent hypergeometric functions. We showed that these complementary functions have curious properties and we gave a graphical demonstration of that. Interestingly, the complementary sine (cosine) series is an even (odd) function in space and satisfy cosine-like (sine-like) boundary condition at the origin. Looking at these functions far away from the origin, one could never predict this interesting and non-trivial behavior. These complementary solutions of the reference wave equation were constructed using the J-matrix method of scattering. In the Appendix, we showed that a similar phenomenon exists in three dimensions. In that case, the regular spherical Bessel function, which is asymptotically sinusoidal, is written as a series of products of Laguerre and Gegenbauer polynomials. The complementary solution, which is also asymptotically sinusoidal and carries the phase shift, is represented by a series of products of the Laguerre polynomials and the hypergeometric functions. In all cases, eliminating the lowest few terms in the series will result in vanishingly small values of these functions within a finite region near the origin. The size of the region increases with the number of eliminated terms.

## Appendix: The three-dimensional case

The reference wave equation for spherically symmetric interaction in 3D, which is analogous to Eq. (2) in 1D, reads as follows

$$\left[\frac{d^2}{dy^2} - \frac{\ell(\ell+1)}{y^2} + \mu^2\right]\psi^\ell(y,\mu) = 0, \quad y \in \mathbb{R}^+, \tag{16}$$

where $y = \lambda r$, $r$ is the radial coordinate, and $\ell$ is the angular momentum quantum number. Here, the integration measure is $\int_0^\infty ... \, dy$. The two independent real solutions of this reference wave equation for $E > 0$ are well-known [2]. They are written in terms of the spherical Bessel and Neumann functions as follows:

$$\psi_{reg}^\ell(y,\mu) = \tfrac{2A}{\sqrt{\pi}}(\mu y) j_\ell(\mu y), \text{ and} \tag{17a}$$

$$\psi_{irr}^\ell(y,\mu) = \tfrac{2B}{\sqrt{\pi}}(\mu y) n_\ell(\mu y), \tag{17b}$$

where $A$ and $B$ are arbitrary real constants. The first is regular (at the origin) and energy-normalized as $\langle\psi_{reg}^\ell | \psi_{reg}^{\prime\ell}\rangle = A\,\delta(\mu-\mu')$. For $\ell > 0$, the second is irregular and is not square integrable. Near the origin they behave as $\psi_{reg}^\ell \to y^{\ell+1}$ and $\psi_{irr}^\ell \to y^{-\ell}$. On the other hand, asymptotically ($y \to \infty$) they are sinusoidal: $\psi_{reg}^\ell \to \tfrac{2A}{\sqrt{\pi}}\sin(\mu y - \pi\ell/2)$ and $\psi_{irr}^\ell \to \tfrac{2B}{\sqrt{\pi}}\cos(\mu y - \pi\ell/2)$. Therefore, unlike the 1D case, only one of the two independent solutions of the reference wave equation in 3D is regular for all $\ell$. We write it as $f_\mu^\ell(y) = \psi_{reg}^\ell(y,\mu)$. A complete $L^2$ basis set that is compatible with this regular solution and with the 3D problem has the following elements [7]

$$\phi_n^\ell(y) = a_n y^{\ell+1} e^{-y/2} L_n^{2\ell+1}(y), \tag{18}$$

where $L_n^{2\ell+1}(y)$ is the associated Laguerre polynomial of order $n$ and the normalization constant is $a_n = \sqrt{n!/(n+2\ell+1)!}$. Therefore, expanding the regular solution in this $L^2$ basis as $f_\mu^\ell(y) = \tfrac{2A}{\sqrt{\pi}}(\mu y) j_\ell(\mu y) = \sum_{n=0}^\infty s_n^\ell(\mu)\phi_n^\ell(y)$ and using the orthogonality property of the Laguerre polynomials [8], we obtain the following integral representation of the expansion coefficients

$$s_n^\ell(\mu) = A\sqrt{2\mu}\, a_n \int_0^\infty y^{\ell+\frac{1}{2}} e^{-y/2} L_n^{2\ell+1}(y) J_{\ell+\frac{1}{2}}(\mu y)\, dy, \tag{19}$$

where we have used the original Bessel functions in writing $j_\ell(z) = \sqrt{\tfrac{\pi}{2z}} J_{\ell+\frac{1}{2}}(z)$. This integral is evaluated in [9] giving

$$s_n^\ell(\mu) = A\pi^{-\frac{1}{2}} 2^{\ell+1} \Gamma(\ell+1)\, a_n (\sin\theta)^{\ell+1} C_n^{\ell+1}(\cos\theta), \tag{20}$$

where $\cos\theta = \tfrac{\mu^2-1/4}{\mu^2+1/4}$, $0 < \theta \leq \pi$, and $C_n^{\ell+1}(z)$ is the ultra-spherical (Gegenbauer) polynomial. Using the differential equation, recursion relation, and orthogonality property of the Laguerre polynomials [8], we obtain the following tridiagonal matrix representation of the reference wave operator

$$J_{nm}^\ell(\mu) = \langle\phi_n^\ell |\left[\frac{d^2}{dy^2} - \frac{\ell(\ell+1)}{y^2} + \mu^2\right]|\phi_m^\ell\rangle = (\mu^2+1/4)$$
$$\times\left[2(n+\ell+1)\cos\theta\,\delta_{n,m} - \sqrt{n(n+2\ell+1)}\,\delta_{n,m+1} - \sqrt{(n+1)(n+2\ell+2)}\,\delta_{n,m-1}\right] \tag{21}$$

Therefore, the matrix representation of the reference wave equation becomes equivalent to the following three-term recursion relation for the expansion coefficients $s_n^\ell$

$$2(n+\ell+1)\cos\theta\, s_n^\ell = \sqrt{n(n+2\ell+1)}\, s_{n-1}^\ell + \sqrt{(n+1)(n+2\ell+2)}\, s_{n+1}^\ell, \quad n \geq 1, \quad (22a)$$

$$2(\ell+1)\cos\theta\, s_0^\ell = \sqrt{2(\ell+1)}\, s_1^\ell. \quad (22b)$$

Moreover, using the differential equation for the Gegenbauer polynomials [10] we can show that $s_n^\ell(\mu)$ satisfies the following second order differential equation in the energy

$$\left[ (1-x^2)\frac{d^2}{dx^2} - x\frac{d}{dx} - \frac{\ell(\ell+1)}{1-x^2} + (n+\ell+1)^2 \right] s_n^\ell(\mu) = 0, \quad (23)$$

where $x = \cos\theta$.

The complementary solution $g_\mu^\ell(y)$ that we are interested in spans the same solution space and could, therefore, be written as $g_\mu^\ell(y) = \sum_{n=0}^\infty c_n^\ell(\mu)\phi_n^\ell(y)$. The new independent expansion coefficients $c_n^\ell(\mu)$ are required to comply with three conditions that are analogous to those in the 1D case stated above Eq. (12). As a result, one obtains the following [11]

$$c_n^\ell(\mu) = -A\pi^{-1} 2^{\ell+1}\Gamma\left(\ell+\tfrac{1}{2}\right) a_n (\sin\theta)^{-\ell}\, {}_2F_1\left(-n-2\ell-1, n+1; -\ell+\tfrac{1}{2}; \sin^2\tfrac{\theta}{2}\right), \quad (24)$$

where ${}_2F_1(a,b;c;z)$ is the hypergeometric function. Figure 3a and Fig. 3b show $f_\mu^\ell(y)$ and $g_\mu^\ell(y)$ for $\mu = 1$ and for $\ell = 0$ and $\ell = 3$, respectively. Figure 4a is a reproduction of Fig. 3a but after eliminating the lowest 30 terms in the series. The functions become very small near the origin but maintain their magnitude and sinusoidal behavior far away from the origin (Fig. 4b). Finally, it is worthwhile noting that there exists another basis for the solution space of the reference problem in 3D, which carries a similar interesting representation. It is referred to as the "oscillator" basis whose elements are: $\phi_n^\ell(y) = \sqrt{\frac{2\Gamma(n+1)}{\Gamma(n+\ell+3/2)}}\, y^{\ell+1} e^{-y^2/2} L_n^{\ell+\frac{1}{2}}(y^2)$ [12].

**Figure Captions:**

**Fig. 1:** Plot of the regular cosine/sine solutions $f_\mu^\pm(y)$ (red curve) and the irregular sine/cosine solutions $g_\mu^\pm(y)$ (blue curve) in 1D for $\mu = 1.2$. We set $A = B = 1$ in arbitrary units.

**Fig. 2:** Plot of $f_\mu^+(y)$ (red curve) and $g_\mu^+(y)$ (blue curve) for $\mu = 1.2$ but after eliminating the lowest $N$ terms in their series representation. We took $N = 10$ in (a) and $N = 20$ in (b) and set $A = B = 1$ in arbitrary units.

**Fig. 3:** Plot of the regular solution $f_\mu^\ell(y)$ (red curve) and the irregular solution $g_\mu^\ell(y)$ (blue curve) in 3D for $\mu = 1$. We took $\ell = 0$ in (a) and $\ell = 3$ in (b) and set $A = B = 1$ in arbitrary units.

**Fig. 4:** (a) Plot of $f_\mu^\ell(y)$ (red curve) and $g_\mu^\ell(y)$ (blue curve) in 3D for $\mu = 1$ and $\ell = 0$ after eliminating the lowest 30 terms in their series representation, (b) The same but far from the origin.

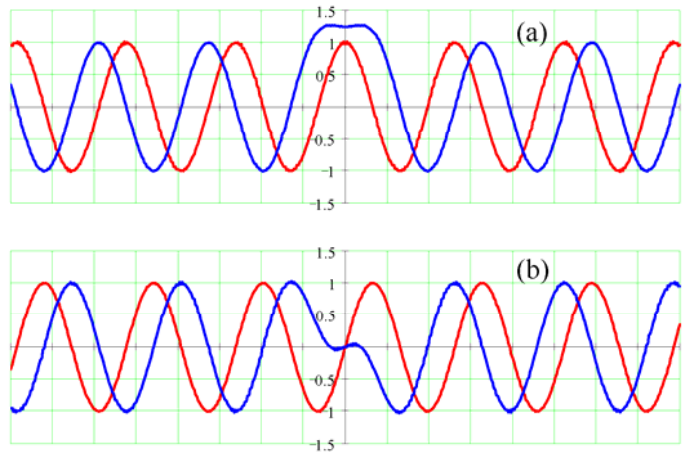

**Fig. 1**

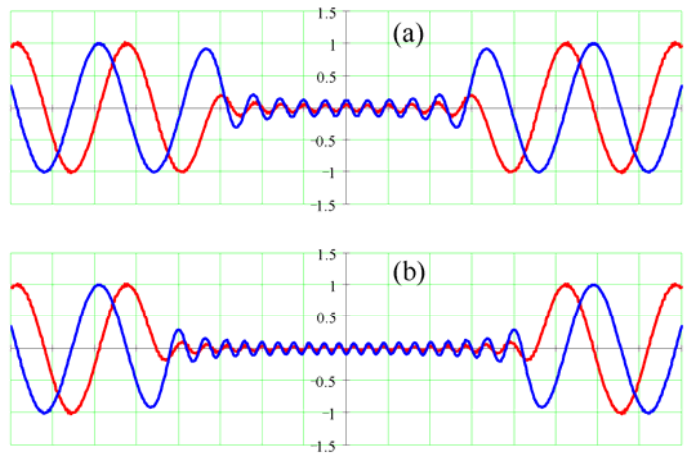

**Fig. 2**

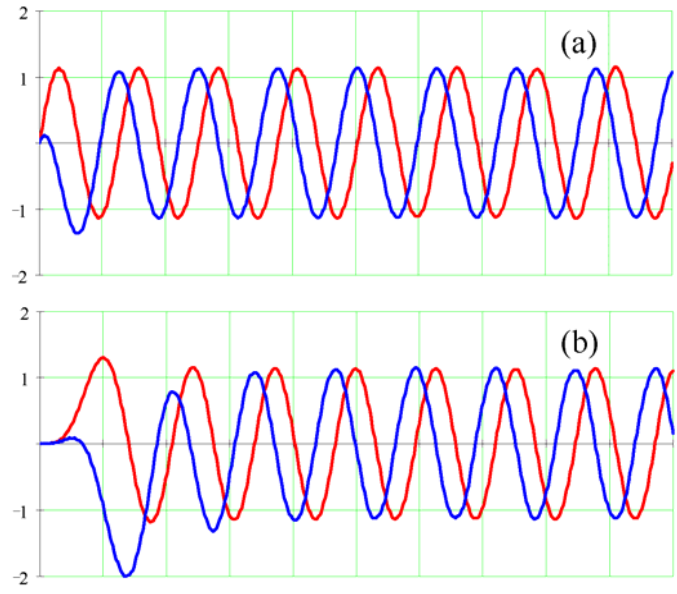

**Fig. 3**

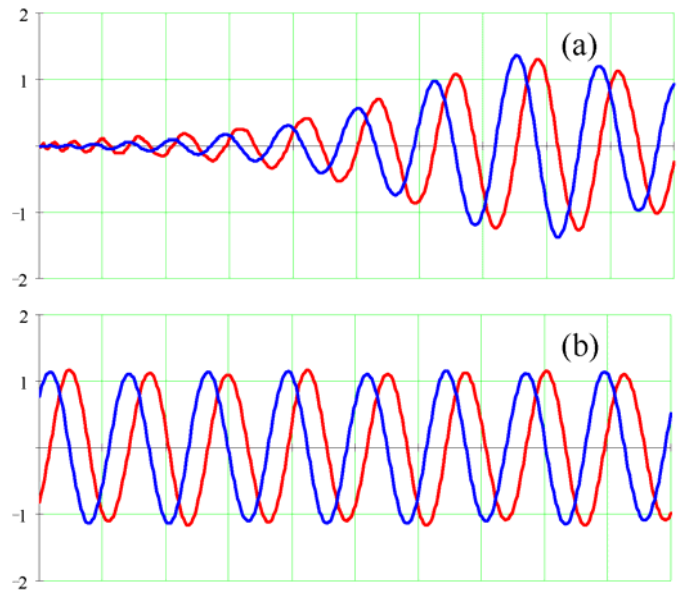

**Fig. 4**